# Total Differential Errors in Two-Port Network Analyser Measurements

N.I. Yannopoulou and P.E. Zimourtopoulos

Since S-parameter measurements without uncertainty cannot claim any credibility, the uncertainties in full two-port Vector Network Analyser (VNA) measurements were estimated using total complex differentials (Total Differential Errors). To express precisely a comparison relation between complex differential errors, their differential error regions (DERs) were used. To demonstrate the method in the most accurate case of a direct zero-length thru, practical results are presented for commonly used Z-parameters of a simple, two-port, DC resistive T-network, which was built and tested against frequency with a VNA measurement system extended by two lengthy transmission lines.

*Introduction*: It is well known that in full two-port VNA measurements the S-parameters for a two-port Device Under Test (DUT) are given in terms of their 4 measurements $m_{ij}$, i=1,2, j=1,2 by

$$S_{11} = \{[(m_{11} - D)/R][1 + (m_{22} - D')M'/R'] - L(m_{21} - X)(m_{12} - X')/(TT')\}/H \qquad (1)$$

$$S_{21} = \{[1 + (m_{22} - D')(M'-L)/R'](m_{21} - X)/T\}/H \qquad (2)$$

$$H = [1 + (m_{11} - D)M/R][1 + (m_{22} - D')M'/R'] - LL'(m_{21} - X)(m_{12} - X')/(TT') \qquad (3)$$

$S_{22}$, $S_{12}$ have expressions that result from (1)-(2) by substituting i,j with j,i and D, M, R, L, T, X with D', M', R', T', L', X' and vice-versa [1]. These 12 quantities have been defined as system errors [2]. Stumper gave non-generalised expressions for the partial deviations of S-parameters due to calibration standard uncertainties, in 2003 [3]. Furthermore, the developed total differential errors for full one-port VNA measurements [4] are also not generalised in the two-port case. To the best of the authors' knowledge, there are no analytical expressions for total differential errors in full two-port VNA measurements.

*Theory*: Since S-parameters are functions of 16 complex variables, their total differential errors were initially expressed as

$$\begin{aligned}
dS_{11} = \{& T T'(1 - MS_{11})[R' + M'(m_{22} - D')](dm_{11} - dD) \\
& - RR'L(1 - L'S_{11})[(m_{21} - X)(dm_{12} - dX') + (m_{12} - X')(dm_{21} - dX)] \\
& + M'T T'[(m_{11} - D)(1 - MS_{11}) - RS_{11}](dm_{22} - dD') \\
& - T T'S_{11}(m_{11} - D)[R' + M'(m_{22} - D')]dM \\
& + T T'(m_{22} - D')[(m_{11} - D)(1 - MS_{11}) - RS_{11}]dM' \\
& - (R'L(1 - L'S_{11})(m_{12} - X')(m_{21} - X) + T T'S_{11}[R' + M'(m_{22} - D')])dR \\
& - (RL(1 - L'S_{11})(m_{12} - X')(m_{21} - X) - T T'[(m_{11} - D)(1 - MS_{11}) - RS_{11}])dR' \\
& - RR'(m_{12} - X')(m_{21} - X)[(1 - L'S_{11})dL - LS_{11}dL'] \\
& + [(m_{11} - D)(1 - MS_{11}) - RS_{11}][R' + M'(m_{22} - D')](T'dT + TdT')\}/P \qquad (4)
\end{aligned}$$

$$dS_{21} = \{- MT\, T'S_{21}[R' + M'(m_{22} - D')](dm_{11} - dD) + RR'LL'S_{21}(m_{21} - X)(dm_{12} - dX')$$

$$+ R\{T'[R' + (m_{22} - D')(M' - L)] + R'LL'S_{21}(m_{12} - X')\}(dm_{21} - dX)$$

$$+ T'(R(m_{21} - X)(M' - L) - M'TS_{21}[R + M(m_{11} - D)])(dm_{22} - dD')$$

$$- T\, T'S_{21}(m_{11} - D)[R' + M'(m_{22} - D')]dM$$

$$+ T'(m_{22} - D')(R(m_{21} - X) - TS_{21}[R + M(m_{11} - D)])dM'$$

$$+ \{(m_{21} - X)(T'(m_{22} - D')(M' - L) + R'[T' + LL'S_{21}(m_{12} - X')])$$

$$- T\, T'S_{21}[R' + M'(m_{22} - D')]\}dR$$

$$+ (R(m_{21} - X)[T' + LL'S_{21}(m_{12} - X')] - T\, T'S_{21}[R + M(m_{11} - D)])dR'$$

$$+ R(m_{21} - X)[R'L'S_{21}(m_{12} - X') - T'(m_{22} - D')]dL$$

$$+ RR'LS_{21}(m_{12} - X')(m_{21} - X)dL' - T'S_{21}[R + M(m_{11} - D)][R' + M'(m_{22} - D')]dT$$

$$+ (R(m_{21} - X)[R' + (m_{22} - D')(M' - L)]$$

$$- TS_{21}[R + M(m_{11} - D)][R' + M'(m_{22} - D')])dT'\}/P \tag{5}$$

$$P = T\, T'[R' + M'(m_{22} - D')][R + M(m_{11} - D)] - RR'LL'(m_{12} - X')(m_{21} - X) \tag{6}$$

$dS_{22}$ and $dS_{12}$ resulted from (4), (5) with the mentioned substitutions. X, X' errors stand for crosstalk measurements. D, M, R (D', M', R') errors are uniquely determined in terms of 3 standard loads A, B, C (A', B', C') and their 3 measurements a, b, c (a', b', c'), by full one-port VNA measurements, so the number of independent complex variables increases from 16 to 22. L, T (L', T') errors are accurately determined after the replacement of DUT with a direct thru (or approximately, if an adapter is used instead) in terms of new measurements $t_{11}$, $t_{21}$ ($t_{22}$, $t_{12}$) and of previously found quantities. Their expressions were appropriately stated as

$$L = [\sum (ab + ct_{11})C(B - A)]/E \tag{7}$$

$$T = (t_{21} - X)[\prod (A - B)(a - b)]/(E\, [\sum cC(B - A)]) \tag{8}$$

$$E = \sum (ab + ct_{11})(B - A) \tag{9}$$

where $\Sigma$ and $\Pi$ produce two more terms, from the given one, by cyclic rotation of the letters a, b, c (a', b', c') or A, B, C (A', B', C'). In this way, each S-parameter has as total differential error dS, a sum of 22 differential terms: 16 due to measurement inaccuracies $dm_{ij}$, dX, dX', $dt_{ij}$, da, db, dc, da', db', dc' and 6 due to standard uncertainties given by their manufacturer dA, dB, dC, dA', dB', dC'. The expressions for dD, dM, dR (dD', dM', dR') are known [4]. The expressions for the rest of differential errors were developed as

$$dL = \{\Sigma (B - C)(b - t_{11})(c - t_{11}) [(B - C)(b-a)(c-a)dA - (b - c)(B - A)(C - A)da]$$
$$+ [\Pi (A - B)(a - b)] dt_{11}\}/E^2 \qquad (10)$$

$$dT = \{\Sigma (t_{21} - X)(b - c)(B - C) [(t_{11} - c)(b - a)^2 B(A^2 + C^2) + (b - t_{11})(c - a)^2 C(A^2 + B^2)$$
$$- 2ABC(b - c)(t_{11}(b + c - 2a) - bc + a^2)] [(B - C)(b - a)(c - a)dA$$
$$- (b - c)(B - A)(C - A)da]\}/(E^2 [\Sigma cC(B - A)]^2)$$
$$+ [\Pi (A - B)(a - b)]\{[(t_{21} - X) \Sigma a(B - C)/E]dt_{11} + dt_{21} - dX\}/(E [\Sigma cC(B - A)]) \qquad (11)$$

Each complex differential error defines a Differential Error Region (DER) on the complex plane with projections to coordinate axes the Differential Error Intervals (DEIs) [4]. Obviously, any quantity differentiably dependent on the above variables has also a DER. For example, after another correction to the given S to Z-parameters relations [5], the Z-DERs are resulted from

$$dZ_{11} = 2Z_0[(1 - S_{22})^2 dS_{11} + (1 - S_{22})S_{21}dS_{12} + (1 - S_{22})S_{12}dS_{21} + S_{12}S_{21}dS_{22}]$$
$$/ [(1 - S_{11})(1 - S_{22}) - S_{12}S_{21}]^2 \qquad (12)$$

$$dZ_{21} = 2Z_0[(1 - S_{22})S_{21}dS_{11} + S_{21}^2 dS_{12} + (1 - S_{11})(1 - S_{22})dS_{21}$$
$$+ (1 - S_{11})S_{21}dS_{22}] / [(1 - S_{11})(1 - S_{22}) - S_{12}S_{21}]^2 \qquad (13)$$

while $dZ_{22}$, $dZ_{12}$ result from (12), (13) by application of the mentioned substitutions.

*Results*: Six calibration standards, in pairs of opposite sex, were used and their manufacturers' data were substituted in the developed expressions: $A = -1 = A'$, $0 \leq$

d|A| = d|A'| ≤ 0.01, −180° ≤ d$\varphi_A$ = d$\varphi_{A'}$ ≤ −178° or 178° ≤ d$\varphi_A$ = d$\varphi_{A'}$ ≤ 180°, B = 0 = B', |dB| = 0.029 = |dB'|, C = 1 = C', −0.01 ≤ d|C| = d|C'| ≤ 0 and −2° ≤ d$\varphi_C$ = d$\varphi_{C'}$ ≤ +2°. The inaccuracy of any VNA measurement was conservatively considered as a symmetric interval defined by just 1 unit in the last place of the corresponding mantissa, both in modulus and argument. Consequently, each S-DER is a sum of 20 parallelograms and 2 circles, with a contour of 160 vertices at most [4]. To demonstrate the method, a typical T-network of common resistors with nominal DC values $Z_1$=24.2 Ω, $Z_2$=120 Ω for the horizontal arms and $Z_{12}$=1.1 Ω for the vertical arm, were soldered on type-N base connectors of opposite sex and enclosed in an aluminium box, to form a two-port DUT. The VNA measurement system was extended by two transmission lines of 3.66 m and 14 m, respectively, up to the DUT. The DUT was tested from 2 to 1289 MHz in 13 MHz steps. The frequency 1003 MHz was selected to illustrate the proposed method for S-DERs shown in Fig. 1. To study the total differential error, dS was expressed as dU + dI, where dU is due to the uncertainty of 6 standards and dI to the inaccuracy of 16 measurements. The contribution of these, conservatively considered measurement inaccuracies to the total differential error is as much significant as the uncertainties of standard loads are. For example, computations for $S_{12}$ over the whole measurement band show that max|dU| and max|dI| contribute ~35%-80% and ~25%-70% to max|d$S_{12}$|, respectively. In addition, Fig. 1 shows how the projections of each S-DER result its real and imaginary DEI. To display the variation of S-DER against frequency, a number of selected S-DER frames are shown in Fig. 2 as beads on a space-curved filament. It is worth mentioning that $S_{11}$-DER ($S_{22}$-DER) was greater than it resulted from appropriately organised full one-port measurements, as it was expected. Finally, the computed Z-DEIs are shown in Fig. 3, along with their LF Z-values.

Therefore, the proposed method may be efficiently used to estimate uncertainties in any case where the process equations (1), (2) and (4), (5) can find application.

Authors' affiliations:

N.I. Yannopoulou, P.E. Zimourtopoulos (Antennas Research Group, Department of Electrical and Computer Engineering, Democritus University of Thrace, V.Sofias 12, Xanthi, 671 00, Greece)

E-mail: **yin@antennas.gr**

Figure captions:

Fig. 1 Typical S-DERs at 1003 MHz.

Fig. 2 S-DERs against frequency.

Fig. 3 Z-DEIs against frequency.

Figure1

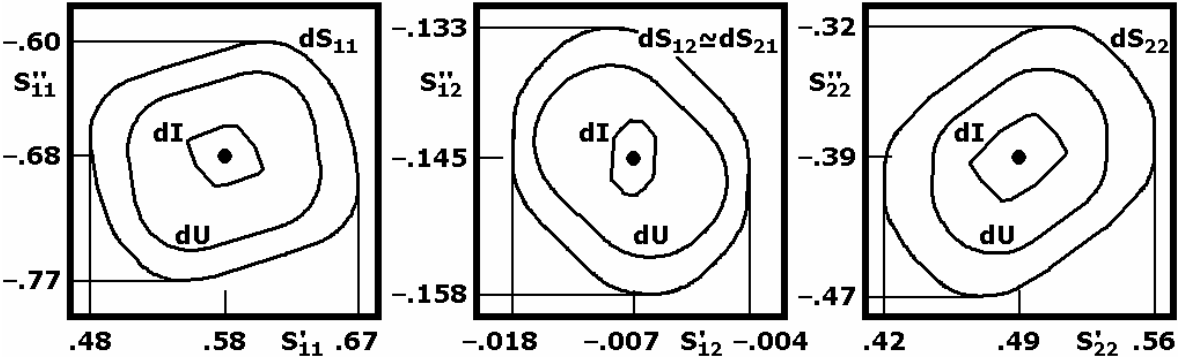

Figure 2

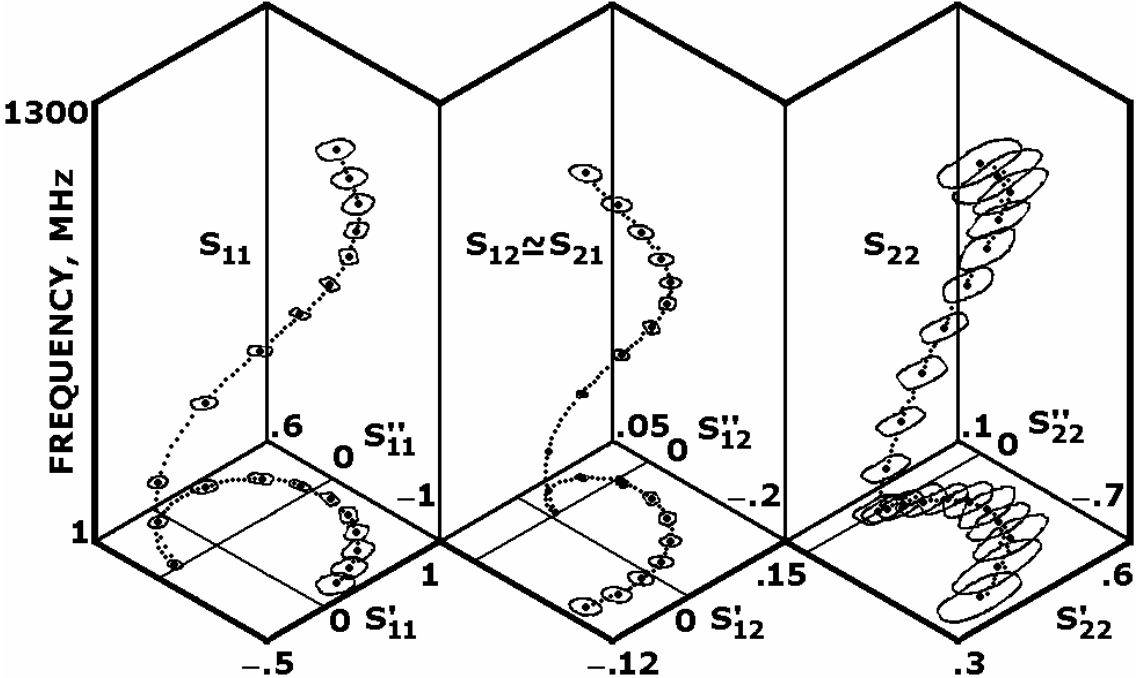

Figure 3

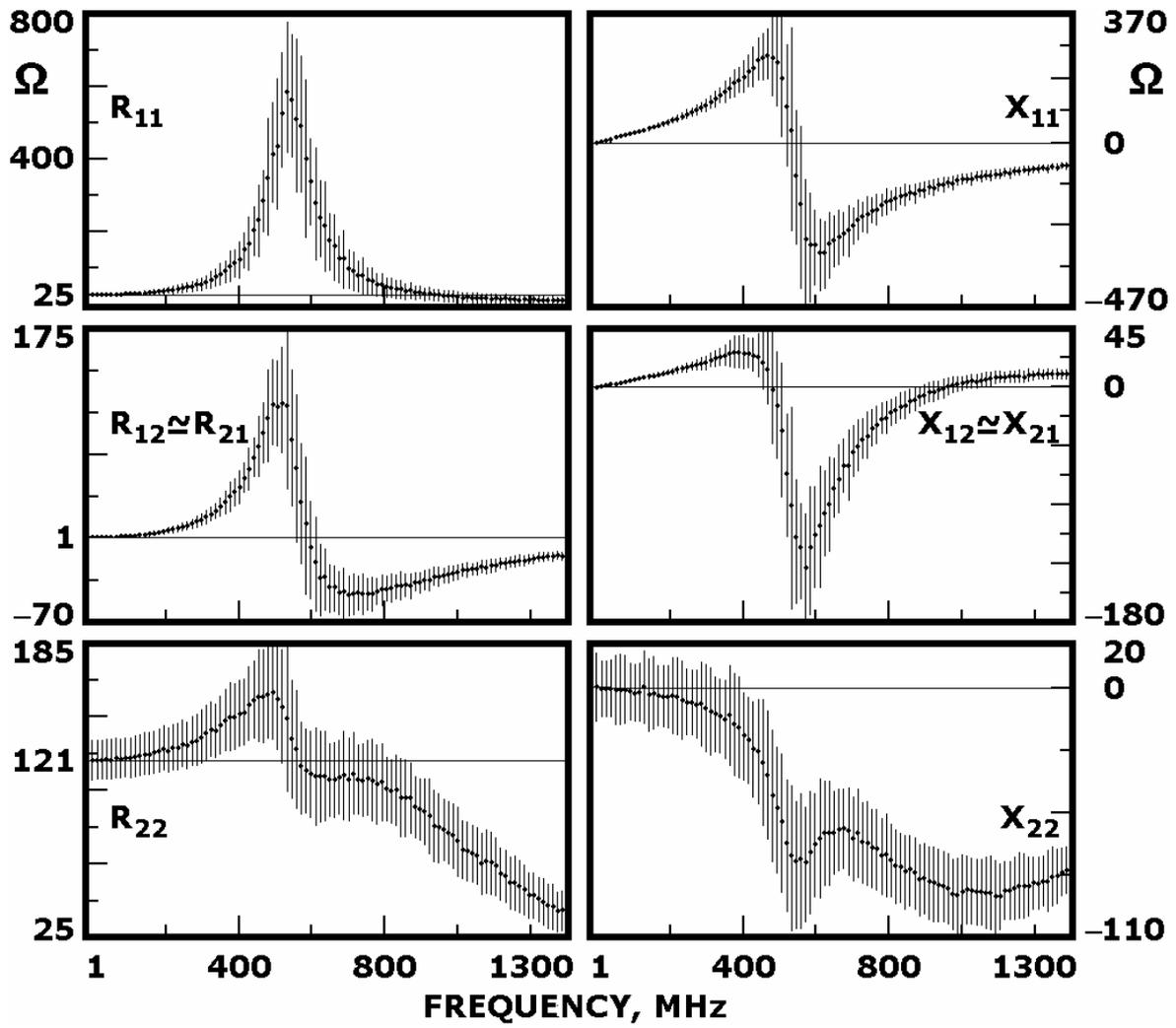